\documentclass[aps,showpacs,pra,superscriptaddress,]{revtex4}
\usepackage{amsmath}
\usepackage{amsfonts}
\usepackage{amssymb}
\usepackage{bm}
\usepackage{graphicx}

\begin{document}

\title{Periodic and localized waves in parabolic-law media with third- and
fourth-order dispersions}
\author{ Houria Triki}
\affiliation{Radiation Physics Laboratory, Department of Physics, Faculty of Sciences,
Badji Mokhtar University, P. O. Box 12, 23000 Annaba, Algeria}
\author{Vladimir I. Kruglov}
\affiliation{Centre for Engineering Quantum Systems, School of Mathematics and Physics,
The University of Queensland, Brisbane, Queensland 4072, Australia}

\begin{abstract}
We study the propagation of femtosecond light pulses inside an optical fiber
medium exhibiting higher-order dispersion and cubic-quintic nonlinearities.
Pulse evolution in such system is governed by a higher-order nonlinear Schr%
\"{o}dinger equation incorporating second-, third-, and fourth-order
dispersions as well as cubic and quintic nonlinearities. Novel classes of
periodic wave solutions are identified for the first time by means of an
appropriate equation method. Results presented indicated the potentially
rich set of periodic waves in the system under the combined influence of
higher-order dispersive effects and cubic-quintic nonlinearity. Solitary
waves of both bright and dark types are also obtained as a limiting case for
appropriate periodic solutions. It is found that the velocity of these
structures is uniquely dependent on all orders of dispersion. Conditions on
the optical fiber parameters for the existence of these stable nonlinear
wave-forms are presented as well.
\end{abstract}

\maketitle
\affiliation{Radiation Physics Laboratory, Department of Physics, Faculty of Sciences,
Badji Mokhtar University, P. O. Box 12, 23000 Annaba, Algeria}
\author{Vladimir I. Kruglov}
\affiliation{Centre for Engineering Quantum Systems, School of Mathematics and Physics,
The University of Queensland, Brisbane, Queensland 4072, Australia}

\section{Introduction}

A soliton in an optical fiber medium can form when the group velocity
dispersion is exactly balanced by self-phase modulation. This localized
pulse is found in two distinct types called bright and dark solitons which
are existent in the anomalous and normal dispersion regimes, respectively.
The unique property of optical solitons, either bright or dark, is their
particle-like behavior in interaction \cite{Kivshar}. Because of their
robust nature, such wave packets have been successfully utilized as the
information carriers (optical bits) to transmit digital signals over long
propagation distances.

Studies of soliton formation in the femtosecond time scale is an important
direction of research in nonlinear optics. This because femtosecond duration
pulses are required for a wide-ranging potential applications such as
ultrahigh-bit-rate optical communication systems, optical sampling systems,
infrared time-resolved spectroscopy, and ultrafast physical processes \cite%
{Ag,Alka}. But when these ultrashort pulses are injected in a fiber medium,
several higher-order nonlinear effects come into play along with dispersive
effects which may significantly change the physical features and stability
of optical soliton propagation. Important higher-order effects include
third-order dispersion, self-steepening, and self-frequency shift which
become important if light pulses are shorter than $100$ fs \cite{Ag}. Taking
into account the influence of various processes appearing in the femtosecond
regime, the description of signal propagation through an optical fiber
medium can be achieved by use of the NLS family of equations incorporating
additional higher-order terms. Compared with solitons in Kerr-like media,
solitary waves supported by higher-order nonlinear and dispersive effects
when they exist can demonstrate much richer dynamics as they propagate
through the system. The contribution of these higher-order effects can also
lead to the formation of novel structures in optical media, including for
example dipole solitons \cite{Am}, W-shaped solitons \cite{ZH}, and
multipole solitons \cite{T2}. So the higher-order effects play a critical
role in the formation of femtosecond solitons.

Recently, attention has been focused on analyzing the dynamic behavior of
soliton pulses in optical fibers exhibiting second-, third-, and
fourth-order dispersions \cite{K1,K2,K3}. In addition to localized pulses,
periodic waves play a significant role in the analysis of the data
transmission in fiber-optic telecommunications links \cite{Dai}. Because of
their structural stability with respect to the small input profile
perturbations and collisions \cite{Petnikova}, this kind of nonlinear waves
serves as a model of pulse train propagation in optics fibers\ \cite{Dai}.
It is relevant to mention that the occurrence of periodic waves is not only
restricted to optical fibers \cite{Chow,Chow1}, but also to other physical
systems such as Bose-Einstein condensates \cite{FKh,Chow2}, nonlinear
negative index materials \cite{Ponz,K4}, and nonlocal media \cite{K5}.

It is of interest to further search for new exact periodic and localized
wave solutions within the higher-order NLSE framework. The exact nature of
the nonlinear waves may be advantageously exploited in designing the optimal
fiber system experiments. Moreover, obtaining such structures is helpful to
recognize various physical phenomena described by the envelope equation. We
should note that exact analytic solutions are also desired for determining
certain important quantities and for simulations. In this paper, we present,
to our knowledge for the first time, a complete map of the existence and
propagation properties of periodic and solitary waves in an optical fiber
medium exhibiting all orders of dispersion up to the fourth order as well as
cubic and quintic nonlinearities. We will introduce a special procedure,
whereby it becomes possible to derive novel periodic and localized wave
solutions of the envelope equation explicitly, and to determine the
conditions under which these structures exist. Importantly, the physically
relevant nonlinear wave solutions presented below indicate the richness of
the fiber medium which is often measured by the variety of nonlinear
structures that it can support. We especially note that the finding new
localized waves is greatly desired as these pulses are ideal instruments for
data transmission over fiber-optic communications lines.

This paper is organized as follows. Section II presents the method used for
obtaining traveling wave solutions of the higher-order NLSE that governs the
propagation of femtosecond light pulses through a highly dispersive optical
cubic-quintic medium. In Sec. III, we identify novel classes of periodic
wave solutions based on an appropriate differential equation. We also find
the solitary wave solutions of the model in the long wave limit and present
the conditions on the optical fiber parameters for their existence. Finally,
we summarize our work in Sec. IV.

\section{Model and traveling waves}

Ultrashort light pulse propagation in a highly dispersive optical fiber
exhibiting a parabolic nonlinearity law obeys the following high dispersive
cubic-quintic NLSE \cite{Palacios}: 
\begin{equation}
i\frac{\partial \psi }{\partial z}=\alpha \frac{\partial ^{2}\psi }{\partial
\tau ^{2}}+i\rho \frac{\partial ^{3}\psi }{\partial \tau ^{3}}-\nu \frac{%
\partial ^{4}\psi }{\partial \tau ^{4}}-\gamma \left\vert \psi \right\vert
^{2}\psi +\mu \left\vert \psi \right\vert ^{4}\psi ,  \label{1}
\end{equation}%
where $\psi (z,\tau )$ is the complex field envelope, $z$ represents the
distance along direction of propagation, and $\tau =t-\beta _{1}z$ is the
retarded time in the frame moving with the group velocity of wave packets.
Also $\alpha =\beta _{2}/2$, $\rho =\beta _{3}/6$, and $\nu =\beta_{4}/24$,
with $\beta _{k}=(d^{k}\beta /d\omega ^{k})_{\omega =\omega _{0}}$ denotes
the k-order dispersion of the optical fiber with $\beta (\omega )$ is the
propagation constant depending on the optical frequency. Parameters $\gamma $
and $\mu $ govern the effects of cubic and quintic nonlinearity,
respectively.

For relatively long optical pulses having width more than $10\mathrm{ps}$,
all three parameters of third- and fourth-order dispersions and quintic
nonlinearity are so small that the model (\ref{1}) reduces to the standard
NLSE which is completely integrable by the inverse scattering method \cite%
{Ag2}. In the absence of quintic nonlinearity ($\mu =0$), solitonlike
solution having a \textrm{sech}$^{2}$ shape and dipole soliton solution of
Eq. (\ref{1}) have been found by employing a regular method \cite{K1,K2}. In
practice, however, the quintic nonlinearity plays a significant role in the
response of many optical materials and can therefore affect the temporal
evolution of optical fields. We therefore analyze the situation in which the
effect of quintic nonlinearity is important and should be taken into account
along with all orders of dispersion up to the fourth order, as described by
the underlying equation (\ref{1}). It is worthy to mention here that optical
media featuring the quintic nonlinearity in the effective index of
refraction include for example chalcogenide glasses \cite{Chen}, organic
polymers \cite{Lawrence}, ferroelectrics \cite{Gu}, and semiconductor-doped
glasses \cite{Roussignol}.

In order to determine the exact traveling wave solutions of Eq. (\ref{1}),
we consider a solution of the form, 
\begin{equation}
\psi (z,\tau )=u(\xi )\exp [i(\kappa z-\delta \tau +\theta )],  \label{2}
\end{equation}%
where $u(\xi )$ is a real amplitude function which depends on the variable $%
\xi =\tau-qz$, with $q=\mathrm{v}^{-1}$ is the inverse velocity. Also the
real parameters $\kappa $ and $\delta $ represent the wave number and
frequency shift respectively, while $\theta $ represents the phase of the
pulse at $z=0$.

From substitution of the representation (\ref{2}) into Eq. (\ref{1}), one
finds the following system of ordinary differential equations, 
\begin{equation}
\nu \frac{d^{4}u}{d\xi ^{4}}-(\alpha +3\rho \delta +6\nu \delta ^{2})\frac{%
d^{2}u}{d\xi ^{2}}+\gamma u^{3}-\mu u^{5}-(\kappa -\alpha \delta ^{2}-\rho
\delta ^{3}-\nu \delta ^{4})u=0,~~~~~~~~~~~~~~~~  \label{3}
\end{equation}%
\begin{equation}
(\rho +4\nu \delta )\frac{d^{3}u}{d\xi ^{3}}+(q-2\alpha \delta -3\rho \delta
^{2}-4\nu \delta ^{3})\frac{du}{d\xi }=0.  \label{4}
\end{equation}
Then from Eq. (\ref{4}), we find that nontrivial solutions for Eqs. (\ref{3}%
) and (\ref{4}) with $\nu \neq 0$ can exist for values of the frequency
shift $\delta $ and inverse velocity $q$ satisfying the relations: 
\begin{equation}
\delta =-\frac{\rho }{4\nu },~~~~q=2\alpha \delta +3\rho \delta ^{2}+4\nu
\delta ^{3}.  \label{5}
\end{equation}%
We can make use of the parameters in (\ref{5}) to determine the wave
velocity $\mathrm{v}=q^{-1}$ as 
\begin{equation}
\mathrm{v}=\frac{8\nu ^{2}}{\rho (\rho ^{2}-4\alpha \nu )}.  \label{6}
\end{equation}%
Relation (\ref{6}) shows that the velocity of propagating waves is uniquely
dependent on the parameters of second-, third-, and fourth-order dispersions
and it does not depend upon the nonlinearity parameters. Therefore, a
natural way to control the velocity of a pulse is to vary various dispersion
parameters in the fiber.

Further substitution of Eq. (\ref{5}) into Eq. (\ref{3}), we obtain an
evolution equation for $u(\xi )$ as 
\begin{equation}
\frac{d^{4}u}{d\xi ^{4}}+\lambda _{0}\frac{d^{2}u}{d\xi ^{2}}+\lambda
_{1}u+\lambda _{2}u^{3}+\lambda _{3}u^{5}=0,~~~~~~~  \label{7}
\end{equation}%
where the parameters $\lambda _{n}$ ($n=0,..,3$) are defined by%
\begin{equation}
\lambda _{0}=\frac{3\rho ^{2}}{8\nu ^{2}}-\frac{\alpha }{\nu },~~~~\lambda
_{1}=-\frac{\kappa }{\nu }-\frac{\rho ^{2}}{16\nu ^{3}}\left( \frac{3\rho
^{2}}{16\nu }-\alpha \right) ,  \label{8}
\end{equation}%
\begin{equation}
\lambda _{2}=\frac{\gamma }{\nu },~~~~\lambda _{3}=-\frac{\mu }{\nu }.
\label{9}
\end{equation}

It is critically important to find exact analytical localized and periodic
solutions of the amplitude equation (\ref{7}) in most general case, when all
parameters of Eq. (\ref{1}) have nonzero values and no constraint for them.
This enables us to examine the individual influence of each type of
dispersive and nonlinear effects on the characteristics of propagating
nonlinear waves. It is interesting to point out that the finding of such
closed form solutions is greatly desired to experiments as they give a
precise formulation of the existing solitary and periodic pulses.

We observe that the nonlinear differential equation (\ref{7}) \ includes two
coexisting cubic $u^{3}$ and quintic $u^{5}$ nonlinear terms in addition to
two even-order derivative terms. In general, it would be very difficult to
find solutions in analytic form for such equation. In the present study, we
have been able to find new types of periodic and localized wave solutions by
using an appropriate equation method. Remarkably, we have found that
integration of Eq. (\ref{7}) leads to physically relevant solutions
satisfying the following equation, 
\begin{equation}
\left( \frac{du}{d\xi }\right) ^{2}=a+bu^{2}+cu^{4},~~~~~~~  \label{10}
\end{equation}%
The corresponding second- and fourth-order differential equations for $u(\xi
)$ read, 
\begin{equation}
\frac{d^{2}u}{d\xi ^{2}}=bu+2cu^{3},~~~~~~~  \label{11}
\end{equation}%
\begin{equation}
\frac{d^{4}u}{d\xi ^{4}}=(b^{2}+12ac)u+20bcu^{3}+24c^{2}u^{5}.~~~~~~~
\label{12}
\end{equation}%
The substitution of Eqs. (\ref{11}) and (\ref{12}) to Eq. (\ref{7}) leads to
the system of algebraic equations as 
\begin{equation}
b^{2}+12ac+\lambda _{0}b+\lambda _{1}=0,~~~~~~~  \label{13}
\end{equation}%
\begin{equation}
20bc+2\lambda _{0}c+\lambda _{2}=0,~~~~24c^{2}+\lambda _{3}=0.~~~~~~~
\label{14}
\end{equation}%
The solution of these algebraic equations yields the parameters for Eq. (\ref%
{10}) in an explicit form as 
\begin{equation}
c=\pm \frac{1}{2}\sqrt{\frac{\mu }{6\nu }},~~~~b=\frac{\alpha }{10\nu }-%
\frac{3\rho ^{2}}{80\nu ^{2}}\mp \frac{\gamma }{10\nu }\sqrt{\frac{6\nu }{%
\mu }},~~~~~~~  \label{15}
\end{equation}%
\begin{equation}
a=\mp \frac{1}{6}\sqrt{\frac{6\nu }{\mu }}(\lambda _{1}+\lambda
_{0}b+b^{2}).~~~~~~~  \label{16}
\end{equation}%
Thus the parameters $c,$ $b$ and $a$ have two different forms with the top
and bottom signs respectively.

We present below a number of novel periodic (or elliptic) solutions of the
model (\ref{1}) based on solving the nonlinear differential equation (\ref%
{10}). These closed form solutions are expressed in terms of Jacobean
elliptic functions of modulus $k$. We further show that special limiting
cases of these families include the bright and dark and solitary wave
solutions.

\section{Periodic and solitary wave solutions}

Before discussing the precise nature of periodic and solitary wave solutions
of the model (\ref{1}), we first consider the transformation of Eq. (\ref{10}%
) based on new function $y(\xi )$ as 
\begin{equation}
u^{2}(\xi )=-\frac{1}{4c}y(\xi ).~~~~~~~  \label{17}
\end{equation}%
Thus, we have found the nonlinear differential equation as 
\begin{equation}
\left( \frac{dy}{d\xi }\right) ^{2}=f(y),~~~~f(y)=\sigma _{1}y+\sigma
_{2}y^{2}-y^{3},~~~~~~~  \label{18}
\end{equation}%
where $\sigma _{1}=-16ac$ and $\sigma _{2}=4b$. The function $f(y)$ can also
be written in the form $f(y)=-y(y-y_{-})(y-y_{+})$ which yields the
nonlinear differential equation, 
\begin{equation}
\left( \frac{dy}{d\xi }\right) ^{2}=-y(y-y_{-})(y-y_{+}).~~~~~~~  \label{19}
\end{equation}%
The polynomial $f(y)$ has tree roots as 
\begin{equation}
y_{0}=0,~~~y_{\pm }=2(b\pm g),~~~~g=\sqrt{b^{2}-4ac}.~~~~~~~  \label{20}
\end{equation}

\begin{description}
\item[\textbf{1. Periodic} $(A+B\mathrm{cn^{2}})^{1/2}$\textbf{-waves}] 
\end{description}

We can order the roots of polynomial $f(y)$ as $y_{1}<y_{2}<y_{3}$ where $%
y_{1}=y_{0}$, $y_{2}=y_{-}$, $y_{3}=y_{+}$. In this case Eqs. (\ref{17}) and
(\ref{18}) yield the periodic solution as 
\begin{equation}
u(\xi )=\pm [ A+B\mathrm{cn^{2}}(w(\xi -\xi _{0}),k)]^{1/2}.~~~~~~~
\label{21}
\end{equation}%
The parameters of this solution are 
\begin{equation}
A=\frac{g-b}{2c},~~~~B=-\frac{g}{c},~~~~~~~  \label{22}
\end{equation}%
\begin{equation}
w=\frac{1}{2}\sqrt{2(b+g)},~~~~k=\sqrt{\frac{2g}{b+g}}.~~~~~~~  \label{23}
\end{equation}%
Here $\mathrm{cn}(w(\xi -\xi _{0}),k)$ is Jacobi elliptic function where the
modulus $k$ belongs the interval $0<k<1$. It follows from this solution the
conditions for parameters as $b>g$, $c<0$ and $b^{2}>4ac$.

The equation for modulus $k$ in Eq. (\ref{23}) yields relation as $%
a=b^{2}(1-k^{2})/c(2-k^{2})^{2}$. Thus, the wave number $\kappa $ by Eqs. (%
\ref{15}) and (\ref{16}) is 
\begin{equation}
\kappa =b\left( \frac{3\rho ^{2}}{8\nu }-\alpha \right) -\frac{\rho ^{2}}{%
16\nu ^{2}}\left( \frac{3\rho ^{2}}{16\nu }-\alpha \right) +\nu b^{2}+\frac{%
12\nu b^{2}(1-k^{2})}{(2-k^{2})^{2}}.  \label{24}
\end{equation}%
Substitution of the solution (\ref{21}) into the wave function (\ref{2})
yields the following family of periodic wave solutions for the high
dispersive cubic-quintic NLSE (\ref{1}):%
\begin{equation}
\psi (z,\tau )=\pm \lbrack A+B\mathrm{cn^{2}}(w(\xi -\xi _{0}),k)]^{1/2}\exp
[i(\kappa z-\delta \tau +\theta )],  \label{25}
\end{equation}%
where modulus $k$ is an arbitrary parameter in the interval $0<k<1$ and $\xi
_{0}$ is the position of pulse at $z=0$. We note that in the limiting cases
with $k=1$ this periodic wave reduces to a bright-type soliton solution.

\begin{figure}[h]
\includegraphics[width=1\textwidth]{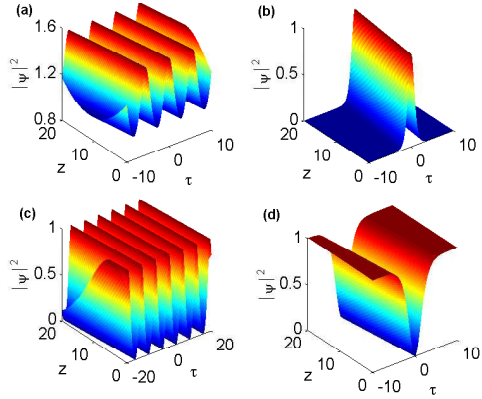}
\caption{Evolution of nonlinear wave solutions (a) $\mathrm{cn}^{2}$ -type
periodic wave solution (\protect\ref{25}) with parameters $\protect\rho %
=0.25,$\ $\protect\alpha =-0.3125,$\ $\protect\nu =0.25,$\ $\protect\gamma %
=-1.5,$ $\protect\mu =0.6144,$ $\protect\xi _{0}=0,$ and\textbf{\ }$k=0.6$
(b) bright solitary wave (\protect\ref{28}) with parameters $\protect\alpha %
=0.4,$ $\protect\rho =1,$ $\protect\nu =0.5,$ $\protect\gamma =0.8,$ $%
\protect\mu =0.75,$ \ $\protect\xi _{0}=0$ (c) $\mathrm{cn}$-type periodic
wave (\protect\ref{32}) with parameters $\protect\alpha =0.25,$ $\protect%
\rho =0.5,$ $\protect\nu =-0.5,$ $\protect\gamma =-0.56875,$ $\protect\mu %
=-0.75,$\ $\protect\xi _{0}=0$ (d) dark solitary wave (\protect\ref{36})
with parameters $\protect\alpha =0.4,$ $\protect\rho =1,$ $\protect\nu =0.5,$
$\protect\gamma =1.075,$ $\protect\mu =0.75,$\ $\protect\xi _{0}=0.$}
\label{FIG.1.}
\end{figure}
Figure 1(a) presents the evolution of the $\mathrm{cn^{2}}$-type periodic
wave (\ref{24}) for the physical parameter values $\rho =0.25,$\ $\alpha
=-0.3125,$\ $\nu =0.25,$\ $\gamma =-1.5,$ and $\mu =0.6144$. To satisfy the
parametric conditions $c<0$ and $b^{2}>4ac$, we considered the case of lower
sign in all the parameters given in Eqs. (\ref{15}) and (\ref{16}). Also,
the value of the elliptic modulus $k$ is taken as\textbf{\ }$k=0.6$\textbf{. 
}As concerns the inverse group velocity $q=\mathrm{v}^{-1}$\ of this $%
\mathrm{cn^{2}}$-type periodic wave, it can be determined from the relation (%
\ref{6}) as $q=0.1875.$\ Additionally, the position $\xi _{0}$ of the
periodic waves at $z=0$ is chosen to be equal to zero. As is seen from this
figure, the intensity profile presents an oscillating character which makes
the wave a setting of light pulse train propagation in optical fibers. A
particularly interesting property of this type of periodic waves is that its
oscillating behaviour is superimposed at a nonzero background, which is
advantageous for a wide range of practical applications.

\begin{description}
\item[\textbf{2. Bright solitary waves}] 
\end{description}

We consider the limiting case of solution in Eq. (\ref{21}) with $k=1$.
Thus, we have the soliton solution of Eqs. (\ref{17}) and (\ref{18}) as 
\begin{equation}
u(\xi )=\pm \left( -\frac{b}{c}\right) ^{1/2}\mathrm{sech}(\sqrt{b}(\xi
-\xi_{0})).~~~~~~~  \label{26}
\end{equation}%
The condition $k=1$ in Eq. (\ref{24}) leads to the wave number $\kappa $ as 
\begin{equation}
\kappa =b\left( \frac{3\rho ^{2}}{8\nu }-\alpha \right) -\frac{\rho ^{2}}{%
16\nu ^{2}}\left( \frac{3\rho ^{2}}{16\nu }-\alpha \right) +\nu b^{2}.
\label{27}
\end{equation}
Thus, the bright solitary wave solution can be obtained for the high
dispersive cubic-quintic NLSE (\ref{1}) using Eqs. (\ref{2}) and (\ref{26})
as 
\begin{equation}
\psi (z,\tau )=\pm \left( -\frac{b}{c}\right) ^{1/2}~\mathrm{sech}(\sqrt{b}%
(\xi -\xi_{0}))\exp [i(\kappa z-\delta \tau +\theta )],  \label{28}
\end{equation}
with $b>0$ and $c<0$.

Figure 1(b)\ depicts the evolution of the intensity wave profile of the
solitary wave solution (\ref{27}) for the parameter values\ $\alpha =0.4,$ $%
\rho =1,$ $\nu =0.5,$ $\gamma =0.8,$ $\mu =0.75,$\ and $\xi _{0}=0.$ We have
also considered the case of lower\ sign in Eqs. (\ref{15}) and (\ref{16})
for the condition $c<0$ to be fulfilled. It is interesting to see that this
localized pulse exhibits a $\mathrm{sech}$-type field profile like the
traditional bright solitons of Kerr media, whereas their existence is due to
a balance among higher-order effects of different nature.

\begin{description}
\item[\textbf{3. Periodic} $\mathrm{cn}$\textbf{-waves}] 
\end{description}

We can order the roots of polynomial $f(y)$ as $y_{1}<y_{2}<y_{3}$ where $%
y_{1}=y_{-}$, $y_{2}=y_{0}$, $y_{3}=y_{+}$. In this case Eqs. (\ref{17}) and
(\ref{18}) yield the periodic solution as 
\begin{equation}
u(\xi )=\pm \Lambda \mathrm{cn}(w(\xi -\xi _{0}),k),~~~~~~~  \label{29}
\end{equation}%
where the modulus $k\in (0,1).$ Here $\Lambda $ and $w$ are real parameters
given by 
\begin{equation}
\Lambda =\left( -\frac{(b+g)}{2c}\right) ^{1/2},~~~~w=\sqrt{g},~~~~k=\sqrt{%
\frac{b+g}{2g}}.~~~~~~~  \label{30}
\end{equation}

The equation for modulus $k$ in Eq. (\ref{30}) yields relation as $%
a=b^{2}k^{2}(k^{2}-1)/c(2k^{2}-1)^{2}$. Thus, the wave number $\kappa $ by
Eqs. (\ref{15}) and (\ref{16}) is 
\begin{equation}
\kappa =b\left( \frac{3\rho ^{2}}{8\nu }-\alpha \right) -\frac{\rho ^{2}}{%
16\nu ^{2}}\left( \frac{3\rho ^{2}}{16\nu }-\alpha \right) +\nu b^{2}+\frac{%
12\nu b^{2}k^{2}(k^{2}-1)}{(2k^{2}-1)^{2}}.  \label{31}
\end{equation}%
Substitution of the solution (\ref{29}) into the wave function (\ref{2})
yields the following family of periodic wave solutions for the high
dispersive cubic-quintic NLSE (\ref{1}):%
\begin{equation}
\psi (z,\tau )=\pm \Lambda \mathrm{cn}(w(\xi -\xi _{0}),k)\exp [i(\kappa
z-\delta \tau +\theta )],  \label{32}
\end{equation}%
where modulus $k$ is an arbitrary parameter in the interval $0<k<1$. It
follows from this solution the conditions for parameters as $b+g>0$, $g>b$, $%
c<0$ and $b^{2}>4ac$. In the limiting case with $k=1$ this solution reduces
to the soliton solution given by Eq. (\ref{28}).

In Fig. 1(c), we have shown the evolution of \textrm{cn}-type periodic wave
solution (\ref{32}) for the parameter values $\alpha =0.25,$ $\rho =0.5,$ $%
\nu =-0.5,$ $\gamma =-0.56875,$ $\mu =-0.75,$\ $\xi _{0}=0.$ To satisfy the
condition $c<0$, we have considered the case of lower\ sign in Eqs. (\ref{15}%
) and (\ref{16}). Unlike the preceding $\mathrm{cn^{2}}$-type periodic wave,
the periodic wave in present case propagates on a zero background.

\begin{description}
\item[\textbf{4. Dark solitary} $\mathrm{tanh}$\textbf{-waves}] 
\end{description}

In the case with $g=0$ or $b^{2}=4ac$ we can order the roots of polynomial $%
f(y)$ as $y_{1}=y_{2}<y_{3}$ where $y_{1}=y_{-}$, $y_{2}=y_{+}$, $%
y_{3}=y_{0} $ . Note that for this case we have the condition $b<0$. Thus,
the solution of Eqs. (\ref{17}) and (\ref{18}) have the kink wave solution
as 
\begin{equation}
u(\xi )=\pm \Lambda \mathrm{tanh}(w(\xi -\xi _{0})).~~~~~~~  \label{33}
\end{equation}
The parameters of this solution are 
\begin{equation}
\Lambda =\left( -\frac{b}{2c}\right) ^{1/2},~~~~w=\frac{1}{2}\sqrt{-2b}%
,~~~~~~~  \label{34}
\end{equation}%
where $b<0$ and $c>0$. The condition $b^{2}=4ac$ yields the wave number $%
\kappa $ by Eqs. (\ref{15}) and (\ref{16}) as 
\begin{equation}
\kappa =b\left( \frac{3\rho ^{2}}{8\nu }-\alpha \right) -\frac{\rho ^{2}}{%
16\nu ^{2}}\left( \frac{3\rho ^{2}}{16\nu }-\alpha \right) +4\nu b^{2}.
\label{35}
\end{equation}%
Hence, one obtains a kink solution for Eq. (\ref{1}) of the form%
\begin{equation}
\psi (z,\tau )=\pm \Lambda \mathrm{tanh}(w(\xi -\xi _{0})) \exp [i(\kappa
z-\delta \tau +\theta )].  \label{36}
\end{equation}
Note that this kink solution has the form of dark soliton for intensity $%
I=|\psi (z,\tau )|^{2}=\Lambda^{2} \mathrm{tanh}^{2} (w(\xi -\xi _{0}))$.

Figure 1(d) displays the intensity profile of the solitary wave solution (%
\ref{36}) for the parameter values\ $\alpha =0.4,$ $\rho =1,$ $\nu =0.5,$ $%
\gamma =1.075,$ $\mu =0.75,$\ $\xi _{0}=0.$ To satisfy the conditions $b<0$
and $c>0$, we have considered the case of upper\ sign in Eqs. (\ref{15}) and
(\ref{16}).

\begin{description}
\item[\textbf{5. Periodic} $\mathrm{sn/}\left( 1+\mathrm{cn}\right) $\textbf{%
-waves}] 
\end{description}

We have also found an unbounded periodic solution of Eq. (\ref{10}) of the
form, 
\begin{equation}
u(\xi )=\pm\frac{A\mathrm{sn}\left( w(\xi -\xi _{0}),k\right) }{1+\mathrm{cn}%
\left( w(\xi -\xi _{0}),k\right) }.  \label{37}
\end{equation}%
The parameters for this periodic solution are 
\begin{equation}
A=\sqrt{\frac{b}{2c(1-2k^{2})}},\qquad w=\sqrt{\frac{2b}{1-2k^{2}}}.\qquad
\label{38}
\end{equation}%
This solution takes place for condition $a=b^{2}/4c(1-2k^{2})^{2}$ which
yields the wave number $\kappa $ by Eqs. (\ref{15}) and (\ref{16}) as 
\begin{equation}
\kappa =b\left( \frac{3\rho ^{2}}{8\nu }-\alpha \right) -\frac{\rho ^{2}}{%
16\nu ^{2}}\left( \frac{3\rho ^{2}}{16\nu }-\alpha \right) +\nu b^{2}+\frac{%
3\nu b^{2}}{(1-2k^{2})^{2}}.  \label{39}
\end{equation}

Now, taking into account the representation (\ref{2}), the higher-order NLSE
(\ref{1}) has the following periodic wave solution:%
\begin{equation}
\psi (z,\tau )=\pm\frac{A\mathrm{sn}\left( w(\xi -\xi _{0}),k\right) }{1+%
\mathrm{cn} \left( w(\xi -\xi _{0}),k\right) }\exp [i(\kappa z-\delta \tau
+\theta )],  \label{40}
\end{equation}
where modulus $k$ is an arbitrary parameter for intervals $0< k<1/\sqrt{2}$
and $1/\sqrt{2}<k<1$.

\begin{description}
\item[\textbf{6. Dark solitary} $\mathrm{tanh}/( 1+\mathrm{sech})$ 
\textbf{-waves}] 
\end{description}

The limit $k\rightarrow 1$ in Eq. (\ref{40}) leads to a solitary wave of the
form, 
\begin{equation}
\psi (z,\tau )=\pm\frac{A_{0}\mathrm{tanh}\left( w_{0}(\xi -\xi _{0})
\right) }{1+\mathrm{sech}\left( w_{0}(\xi -\xi _{0}) \right) }\exp [i(\kappa
z-\delta \tau +\theta )].  \label{41}
\end{equation}%
The parameters for this solitary wave are 
\begin{equation}
A_{0}=\sqrt{-\frac{b}{2c}},\qquad w_{0}=\sqrt{-2b},\qquad  \label{42}
\end{equation}%
with $b<0$ and $c>0$ and $a=b^{2}/4c$. This solitary wave has the form of
dark soliton for intensity $I=|\psi (z,\tau )|^{2}$. The wave number $\kappa 
$ for this solitary wave follows from Eq. (\ref{39}) with $k=1$: 
\begin{equation}
\kappa =b\left( \frac{3\rho ^{2}}{8\nu }-\alpha \right) -\frac{\rho ^{2}}{%
16\nu ^{2}}\left( \frac{3\rho ^{2}}{16\nu }-\alpha \right) +4\nu b^{2}.
\label{43}
\end{equation}

\begin{figure}[h]
\includegraphics[width=1\textwidth]{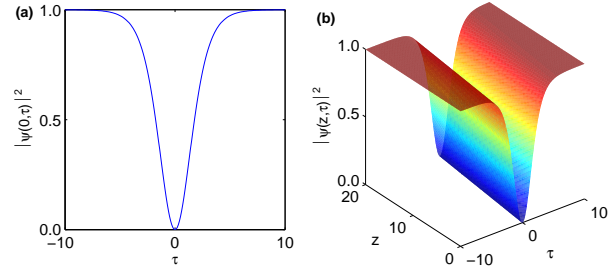}
\caption{(a) Intensity of the solitary wave profile $\left\vert \protect\psi %
(0,\protect\tau )\right\vert ^{2}$ as a function of $\protect\tau $ and its
(b) evolution as computed from Eq. (\protect\ref{41}) for the value $\protect%
\alpha =0.4,$ $\protect\rho =1,$ $\protect\nu =0.5,$ $\protect\gamma =1.075,$
$\protect\mu =0.75,$ and $\protect\xi _{0}=0.$ }
\label{FIG.2.}
\end{figure}
Figure 2(a)\ presents the intensity profile of the optical solitary wave
solution (\ref{41}) for the parameter values $\alpha =0.4,$ $\rho =1,$ $\nu
=0.5,$ $\gamma =1.075,$ $\mu =0.75,$\ $\xi _{0}=0,$ while Fig. 2(b) shows
its evolution. It is interesting to see that this nonlinear waveform is a
dark solitary wave, which can be formed in the fiber medium due to a balance
among all orders of dispersion up to the fourth order and both third- and
fifth-order nonlinearities. Remarkably, the functional form of this solitary
wave is different from the simplest dark solitary wave that has the form 
\textrm{tanh}.

\begin{description}
\item[\textbf{7. Periodic} $\mathrm{cn/}\left( 1+\mathrm{sn}\right) $\textbf{%
-waves}] 
\end{description}

We have found the unbounded periodic solution of Eq. (\ref{10}) of the form, 
\begin{equation}
u(\xi )=\pm\frac{A\mathrm{cn}\left( w(\xi -\xi _{0}),k\right) }{1+\mathrm{sn}%
\left( w(\xi -\xi _{0}),k\right) },  \label{44}
\end{equation}%
where $0\leq k<1$. The parameters for this periodic solution are 
\begin{equation}
A=\sqrt{\frac{b(1-k^{2})}{2c(1+k^{2})}},\qquad w=\sqrt{\frac{2b}{1+k^{2}}}%
,\qquad  \label{45}
\end{equation}%
with $b>0$ and $c>0$. This solution takes place for condition $%
a=b^{2}(1-k^{2})^{2}/4c(1+k^{2})^{2}$ which yields the wave number $\kappa $
by Eqs. (\ref{15}) and (\ref{16}) as 
\begin{equation}
\kappa =b\left( \frac{3\rho ^{2}}{8\nu }-\alpha \right) -\frac{\rho ^{2}}{%
16\nu ^{2}}\left( \frac{3\rho ^{2}}{16\nu }-\alpha \right) +\nu b^{2}+\frac{%
3\nu b^{2}(1-k^{2})^{2}}{(1+k^{2})^{2}}.  \label{46}
\end{equation}%
The substitution of solution (\ref{44}) into Eq. (\ref{2}) yields the family
of periodic bounded solutions for the higher-order NLSE (\ref{1}) of the
form, 
\begin{equation}
\psi (z,\tau )=\pm\frac{A\mathrm{cn}(w(\xi -\xi _{0}) ,k)}{1+\mathrm{sn}%
(w(\xi -\xi _{0}) ,k)}\exp [i(\kappa z-\delta \tau +\theta )],  \label{47}
\end{equation}
where modulus $k$ is an arbitrary parameter in the interval $0\leq k<1$.

\begin{description}
\item[\textbf{8. Periodic} $\mathrm{sn/}\left( 1+\mathrm{dn}\right) $\textbf{%
-waves}] 
\end{description}

We have found the exact periodic bounded solution of Eq. (\ref{10}) of the
form, 
\begin{equation}
u(\xi )=\pm\frac{A\mathrm{sn}\left( w(\xi -\xi _{0}),k\right) }{1+\mathrm{dn}%
\left( w(\xi -\xi _{0}),k\right) }.  \label{48}
\end{equation}%
The parameters for this periodic solution are 
\begin{equation}
A=\sqrt{\frac{bk^{4}}{2c(k^{2}-2)}},\qquad w=\sqrt{\frac{2b}{k^{2}-2}},\qquad
\label{49}
\end{equation}%
where $b<0$ and $c>0$. This solution takes place for condition $%
a=b^{2}k^{4}/4c(k^{2}-2)^{2}$ which yields the wave number $\kappa $ by Eqs.
(\ref{15}) and (\ref{16}) as 
\begin{equation}
\kappa =b\left( \frac{3\rho ^{2}}{8\nu }-\alpha \right) -\frac{\rho ^{2}}{%
16\nu ^{2}}\left( \frac{3\rho ^{2}}{16\nu }-\alpha \right) +\nu b^{2}+ \frac{%
3\nu b^{2}k^{4}}{(k^{2}-2)^{2}}.  \label{50}
\end{equation}%
Thus, the appropriate periodic bounded solutions of Eq. (\ref{1}) are%
\begin{equation}
\psi (z,\tau )=\pm\frac{A\mathrm{sn}(w(\xi -\xi _{0}) ,k)}{1+\mathrm{dn}%
(w(\xi -\xi _{0}),k)}\exp [i(\kappa z-\delta \tau +\theta )],  \label{51}
\end{equation}
where modulus $k$ is an arbitrary parameter in the interval $0<k<1$. It
should be noticed that the limit $k\rightarrow 1$ in this solution yields
the solitary wave solution given in Eq. (\ref{41}).

\begin{figure}[h]
\includegraphics[width=1\textwidth]{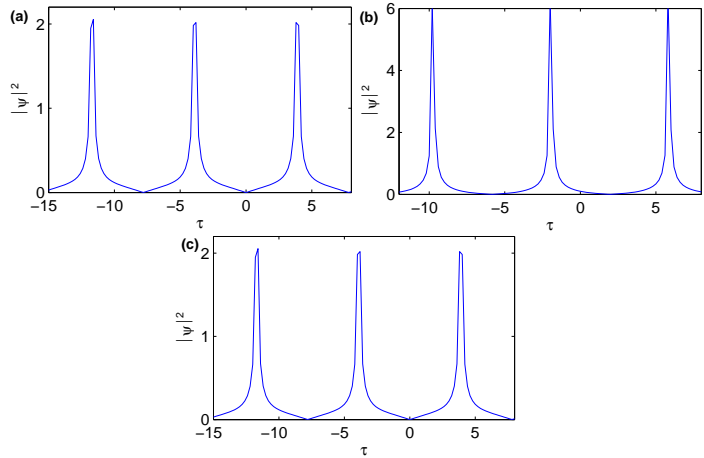}
\caption{Intensity profiles of (a) the periodic wave solution (\protect\ref%
{40}) with parameters $\protect\alpha =1.64,$ $\protect\rho =-0.3,$ $\protect%
\nu =1,$ $\protect\gamma =0.1,$ $\protect\mu =1.5,$ $\protect\xi _{0}=0,$ $%
k=0.6$ (b) the periodic wave solution (\protect\ref{47}) with parameters $%
\protect\alpha =0.305,$ $\protect\rho =-0.2,$ $\protect\nu =0.5,$ $\protect%
\gamma =-1,$ $\protect\mu =0.3072,$ $\protect\xi _{0}=0,$ $k=0.6$ (c) the
periodic wave solution (\protect\ref{51}) with parameters $\protect\alpha %
=0.21,$ $\protect\rho =-0.2,$ $\protect\nu =1.5,$ $\protect\gamma =1,$ $%
\protect\mu =0.0576,$ $\protect\xi _{0}=0,$ $k=0.6$. }
\label{FIG.3.}
\end{figure}
Figure 3(a) depicts a typical example of the time evolution of intensity of
the periodic wave (\ref{40}) for the parameters values $\alpha =1.64,$ $\rho
=-0.3,$ $\nu =1,$ $\gamma =0.1,$ $\mu =1.5$. Then the velocity of the wave
can be determined by using the relation (\ref{6}) as $\mathrm{v}\approx
4.12. $ The results for the unbounded periodic wave (\ref{47}) are
illustrated in Fig. 3(b) for the values $\alpha =0.305,$ $\rho =-0.2,$ $\nu
=0.5,$ $\gamma =-1,$ $\mu =0.3072$. The velocity of this wave can be
calculated with the help of Eq. (\ref{6}) resulting in $\mathrm{v}\approx
17.54$. The intensity profile of the periodic wave solution (\ref{51}) is
shown in Fig. 3(c) for the values $\alpha =0.21,$ $\rho =-0.2,$ $\nu =1.5,$ $%
\gamma =1,$ $\mu =0.0576$. Accordingly, the velocity of the wave is obtained
as $\mathrm{v}\approx 73.77$. Here we considered the case of upper sign in
all the parameters given in Eqs. (\ref{15}) and (\ref{16}) and the initial
position $\xi _{0}$ of the periodic waves is chosen to be equal to zero.%
\textbf{\ }Also, the value of elliptic modulus $k$ is taken as $k=0.6$. We
can see from this figure that the profile of nonlinear waves presents the
periodic property as it propagates through the optical fiber. It is also
interesting to note that the oscillating behaviour of this kind of periodic
waves is superimposed at a zero background.

In view of the above results, we thus see that, in addition to the simplest
periodic waves, novel periodic waves taking the forms (\ref{25}), (\ref{40}%
), (\ref{47}), and (\ref{51}) can also be formed in the fiber medium in the
presence of various higher-order effects. No doubt, this may be helpful for
extending the applicability for periodic wave propagation through highly
dispersive optical fibers. We should note here that the periodic structures
are of increasing interest particularly after the first experimental
observation of the evolution of an arbitrarily shaped input optical pulse
train to the shape preserving Jacobean elliptic pulse-train corresponding to
the Maxwell-Bloch equations \cite{Shultz}. The obtained results also showed
that the presence of higher-order dispersions and quintic nonlinearity leads
to the formation of a novel class of dark-type waves which takes the form
given by Eq. (\ref{40}). Undoubtedly, such ultrashort solitary pulse could
find potential applications in optical communication systems since dark
solitons are more stable against Gordon-Haus jitters in long communication
line, less influenced by noise, and less sensitive to optical fiber loss 
\cite{Yu,Amitava}. We emphasis that periodic and solitary wave solutions
presented in this section are stable to small perturbations. It can be proved using the analytical method developed in Ref. \cite{K2}.

\section{Conclusion}

We have studied the femtosecond light pulse propagation in a highly
dispersive optical fiber governed by a higher-order nonlinear Schr\"{o}%
dinger equation incorporating all orders of dispersion up to the fourth
order as well as cubic and quintic nonlinearities. With use of an
appropriate equation, novel exact periodic wave solutions have been
identified for the model in the presence of various dispersive and nonlinear
effects. Solitary waves have been also obtained which includes both bright
and dark localized solutions. It is found that the velocity of these
structures is uniquely dependent on all orders of dispersion. Moreover, all solutions presented in the paper are stable to small perturbations which follows from appropriate stability analysis. It is apparent that the exact nature of the nonlinear waves presented here can lead to different applications in optical communications.

\end{document}